\documentclass[twocolumn,fullpaper]{jpsj3}

\def\runauthor{Osamu \textsc{Sakai}
}

\newcommand{\Bk}{\mbox{\boldmath $k \:$}}

\newcommand{\Bi}{\mbox{\boldmath $i \:$}}
\newcommand{\Br}{\mbox{\boldmath $r \:$}}

\def\Beqa{\begin{eqnarray}}
\def\Eeqa{\end{eqnarray}}

\def\I{{\rm i}}
\def\VEP{\varepsilon}
\def\P{\partial}

\title{%
Band Calculations for Ce Compounds with AuCu$_{3}$-type Crystal Structure on the  basis of  Dynamical 
Mean Field Theory I. - CePd$_{3}$ and CeRh$_{3}$ - 
}

\author{%
Osamu \textsc{Sakai}\thanks{E-mail address: sakai\_ym@star.ocn.ne.jp}
}

\inst{%
National Institute for Materials Science,
Sengen 1-2-1, Tsukuba 305-0047, Japan \\
}

\recdate{April 19, 2010
}

\abst{%
Band calculations for Ce compounds
with the AuCu$_{3}$-type crystal structure were carried out on the basis of dynamical mean field 
theory (DMFT).
The auxiliary impurity problem was solved by a method named NCA$f^{2}$vc
(noncrossing approximation including the $f^{2}$ state as a vertex correction). 
The calculations take into account the crystal-field splitting, the spin-orbit interaction, 
and the 
correct exchange process of the $f^{1} \rightarrow f^{0},f^{2}$ virtual excitation. 
These are necessary features in the quantitative band theory for Ce compounds 
and
in the calculation of their excitation spectra.
The results of applying the calculation to CePd$_{3}$ and CeRh$_{3}$ are presented as the first in a series of papers. 
The experimental results of the photoemission spectrum (PES), the inverse PES, the angle-resolved PES, and 
the magnetic excitation spectra were reasonably reproduced by the first-principles DMFT band calculation.
At low temperatures, 
the Fermi surface (FS) structure of CePd$_{3}$ is similar to that of the band obtained by the local density approximation.
It gradually changes into a form that is similar to the FS of LaPd$_{3}$ as the temperature increases,
since the $4f$ band shifts to the high-energy side and the lifetime broadening becomes large.
}
\kword{%}
dynamical mean field theory, band theory,
ARPES, magnetic excitation, CePd$_{3}$, CeRh$_{3}$, Cauchy integral
}

\begin{document}
\sloppy
\maketitle

\section{Introduction}
Nonempirical band calculations for strongly correlated electron systems have been extensively developed 
on the basis of dynamical mean field theory (DMFT)~\cite{A1,A2}.
The $4f$ electrons  in Ce compounds are typical strongly correlated electrons~\cite{A3,A4,A5}.
Recently, a DMFT band calculation scheme for Ce compounds was developed in refs.~\citen{A6} and~\citen{A7}. 
In the present work, it is applied to Ce compounds with the AuCu$_{3}$-type crystal structure, which show
a wide variety of $4f$ electronic states from the most itinerant limit to the localized limit.

The $4f$ state splits into the $j=5/2$ ground multiplet and the $j=7/2$ excited
multiplet with a separation of about 0.3~eV owing to the spin-orbit interaction (SOI).
The multiplet shows crystal-field splitting (CFS) of the order of 100~K.
In cubic crystals, the $j=5/2$ multiplet splits into the ($j=5/2$)$\Gamma_{7}$ doublet 
and the ($j=5/2$)$\Gamma_{8}$ quartet.
Hereafter, we call them the $\Gamma_{7}$ and $\Gamma_{8}$ states, respectively.  
It is important to take account of the SOI and CFS effects in $4f$ systems~\cite{A3}.
In DMFT, the correlated band electron problem is mapped onto the calculation of the 
single-particle excitation spectrum 
of the auxiliary impurity Anderson model in an effective medium. 
Reliable methods of solving the impurity Anderson model with CFS and the SOI effect are needed in the DMFT 
band calculation for 
$4f$ compounds.
A theory named  NCA$f^{2}$vc (noncrossing approximation including the $f^{2}$ state as a vertex correction) 
has been developed~\cite{A8,A9,A10,B1} and combined with the linear muffin-tin orbital (LMTO) method~\cite{A11,A12} to carry out the DMFT band 
calculation~\cite{A7}.
NCA$f^{2}$vc can include CFS and the SOI effect, and also the 
correct exchange process of the $f^{1} \rightarrow f^{0}, f^{2}$ virtual excitation. 
The calculation gives an accurate order of the Kondo temperature ($T_{\rm K}$).

The DMFT band calculation will be applied in a series of studies to Ce compounds with the AuCu$_{3}$-type structure: 
CePd$_{3}$, CeRh$_{3}$,
CeIn$_{3}$, and CeSn$_{3}$. 
Each of these materials is classified as a typical example of strongly correlated $4f$ systems~\cite{A5,A13,A14,A15,E1,A16,A17}.

CePd$_{3}$ is a typical heavy Fermion system with $T_{\rm K}$ of about 250~K, 
and has been studied extensively by various methods~\cite{A5,A16,A17,A18,A19,A20,A21,A22,A23,A24,A25,A26,A27,A28,A29}. 
It has a nonmagnetic Fermi liquid (FL) ground state at low temperatures.
A controversy existed in the experimental works on inelastic magnetic excitation, 
but it has recently been resolved by a detailed study of the wave number vector (wave vector) dependence of 
spectra~\cite{A17,A19,A20,A21}.
The wave-vector-integrated-spectrum of the magnetic excitation has a broad peak at approximately 55~meV.
In the single-particle excitation, a broad peak with the $4f$ character was observed on the inverse photoemission spectrum 
(IPES) side (i.e., in the energy region above the Fermi energy ($E_{\rm F}$)). 
This seems to be consistent with the high $T_{\rm K}$ of this compound~\cite{A22,A23,A24,A25}.
However, a strong peak structure has never been observed on the photoemission spectrum (PES) 
side (i.e., in the energy region below $E_{\rm F}$), contradicting other physical properties~\cite{A26,A27}. 
Recent careful study using the $3d$-$4f$ high-resolution resonant photoemission spectrum (RPES) revealed that
the bulk component of the PES of this compound shows strong intensity at $E_{\rm F}$ consistently with high $T_{\rm K}$~\cite{A28}.
The angle-resolved PES (ARPES) has also been studied recently~\cite{A29}. 
It may be worthwhile whether to confirm these recent results of studies are reproduced or not by the first-principles DMFT band calculation. 
CeRh$_{3}$ is known as one of the compounds having the most itinerant 4f states~\cite{A23,A24,A30,A31,A32,A33,A34}.

CeSn$_{3}$ is also known to show the nonmagnetic FL ground state with a high characteristic temperature~\cite{A35,A36}.
CeIn$_{3}$ has an antiferromagnetic ground state with a Neel temperature of $T_{\rm N}=11$~K.
Recently, it was found that  
$T_{\rm N}$ decreases to zero under the pressure of 2.5 GPa, in addition,  
the transition to superconductivity occurs at $T_{\rm SC}=0.2$~K~\cite{A37,A38,A39}.
It will be interesting to study the change of the band structure under pressures by the DMFT calculation.

These compounds commonly have the AuCu$_{3}$-type crystal structure, 
which is classified into the simple cubic ({\it sc}) lattice.
The DMFT band structure of these compounds will be reported in two papers.
In CePd$_{3}$ and CeRh$_{3}$, the hybridization of $4f$ states with $4d$ states of the transition metal is very strong.
The $4d$ states almost sink to below $E_{\rm F}$ in CePd$_{3}$ whereas they are located near $E_{\rm F}$ in
CeRh$_{3}$~\cite{A40}. 
In this paper, calculations for these $4d$ compounds will be reported.
Calculations for CeSn$_{3}$ and CeIn$_{3}$ will be given in a subsequent paper.
Their $4f$ states  hybridize with broad $5p$ states of ligand ions.

Calculated results of CePd$_{3}$ and CeRh$_{3}$ generally show reasonable agreement with experimental results of the PES, IPES, ARPES, 
and inelastic magnetic excitation by neutrons.
However, the calculation gives a higher $T_{\rm K}$ than that expected from experiments when it is examined in detail.
The DMFT band calculation for CePd$_{3}$ gives a Fermi surface (FS) structure similar to that obtained by the local density approximation (LDA) calculation 
at very low temperatures.
When the temperature increases, $4f$ bands shift to the high-energy side and their lifetime broadening increases. 
This leads to the change of the FS structure into one that is similar to the FS of LaPd$_{3}$.
At $T=150$~K, the FS has a different form from both FSs of LaPd$_{3}$ and the LDA band of CePd$_{3}$ 
as an intermediate stage of the change.
The lifetime broadening overcomes the fine wave vector dependence of the $4f$ spectrum at $T=300$~K.
In CeRh$_{3}$, the $4f$ band is located at about 0.9~eV above $E_{\rm F}$, and the dispersion of the DMFT band is almost 
identical to that of the LDA band in the energy region near $E_{\rm F}$.
The density of states (DOS) has an appreciable value slightly below $E_{\rm F}$ in the ARPES.
This low-binding energy part shows a weak wave vector dependence, though no flat bands do not exist in the 
vicinity of the Fermi energy.

In \S 2, we briefly give the formulation on the basis of the LMTO method.
Results of the application to CePd$_{3}$ are shown in \S 3, 
and results for CeRh$_{3}$ are  given in \S 4.
A summary is given in \S 5.
In the appendices, notes on the calculation of the total electron number are given.
An efficient method of calculating the Cauchy integral using the spline interpolation scheme is also presented.
This integral is frequently used in the DMFT calculation.

\section{Formulation}

The method of calculation is described briefly because  
its details have been given in previous papers~\cite{A6,A7}. 
 We consider the excitation spectrum of the following Hamiltonian: 
\Beqa
{\cal H}={\cal H}_{\rm LDA} 
+
\frac{U}{2}\sum_{\Bi}
(\sum_{\Gamma,\gamma}
c^{+}_{\phi^{\rm a}\Bi\Gamma\gamma}
     c_{\phi^{\rm a}\Bi\Gamma\gamma}
 -n^{\rm LDA*}_{\Bi f})^{2}.
\label{eq.Hamiltonian}
\Eeqa
Here, 
$ c_{\phi^{\rm a}\Bi\Gamma\gamma}$ is the annihilation 
operator for the atomic localized state 
$
\phi^{\rm a}_{\Bi\Gamma\gamma}(\Br)
$ 
at site  $\Bi$ with  the 
$\gamma$ orbital of the $\Gamma$-irreducible representation. 
The quantity $n^{\rm LDA*}_{\Bi f}$  
is determined using the 
occupation number of the atomic $4f$ electron per Ce ion in 
the LDA calculation.
We assume that the local Coulomb interaction acts only on the orbital 
$\phi^{\rm a}_{\Bi\Gamma\gamma}$.

The excitation spectrum is expressed by introducing the self-energy
terms~\cite{A6},
\Beqa
{\cal H}_{\rm DMFT}={\cal H}_{\rm LDA} \hspace{5cm} \nonumber \\
+ \sum_{\Bi,(\Gamma,\gamma)}
(\Sigma_{\Gamma}(\VEP+{\rm i}\delta)
 +\VEP^{\rm a}_{\Gamma}-\VEP_{\Gamma}^{\rm LDA})
 |\phi^{\rm a}_{\Bi\Gamma\gamma} >
     < \phi^{\rm a}_{\Bi\Gamma\gamma} |,
\label{eq.DMFT-Hamiltonian}
\Eeqa
where $\VEP^{\rm a}_{\Gamma}$ is the single-electron energy level of 
the $4f$ state, and 
$\VEP_{\Gamma}^{\rm LDA}$ is the energy level 
in the  LDA calculation.
The self-energy $\Sigma_{\Gamma}(\VEP+{\rm i}\delta)$ is calculated by solving the 
auxiliary impurity problem with the use of  
NCA$f^{2}$vc; its outline is described in the Appendix of ref.~\citen{A7}.

In later calculations we will approximate the localized $4f$ state
$\phi^{\rm a}$
by the band center orbital 
$\phi(-)$ 
because its localization is good for the $4f$ state.
$\phi(-)$ has the logarithmic derivative $-\ell-1$ on the muffin-tin surface~\cite{A11,A12}.

In the LMTO method, the Hamiltonian 
${\cal H}_{\rm LDA}$ is diagonalized 
using the LMTO bases,
\Beqa
\psi^{j\Bk}(\Br)
=\sum_{q L}a^{j\Bk}_{q L}
\chi^{\Bk}_{q L}(\Br).
\label{eq.psi}
\Eeqa
Here, $a^{j\Bk}_{q L}$ is the 
expansion coefficient 
of the $j$th eigenvector 
on the LMTO base of the Bloch type, $\chi^{\Bk}_{q L}(\Br)$, with the wave number vector $\Bk$,  
the angular momentum
($\ell,m$), and the spin ($\alpha$) 
at site $q$ 
in the unit cell,
where $L \equiv (\ell,m,\alpha)$~\cite{A12}.

The explicit expression of $\chi^{\Bk}_{q L}(\Br)$ has been given in a 
previous paper~\cite{A6}.
Note that they 
are not orthogonal to each other, but 
the eigenvectors  
$\psi^{j\Bk}$ are orthonormal.

The DMFT band structure is calculated in the following way:
(A) first the LDA part of the Hamiltonian 
${{\cal H}}_{\rm LDA}$ is diagonalized for a given $\Bk$, (B) then the matrix equation of the Greenian is prepared in the 
manifold of the eigenvectors 
$\psi^{j\Bk}$.
The Greenian equation for the given $\Bk$ is written as 
\Beqa
[ zI-D_{\rm LDA}(\Bk)-\Sigma(z)]G(z;\Bk) = I,
\label{eq.G-eq}
\Eeqa
where $I$ is the unit matrix and 
$D_{\rm LDA}(\Bk)$ is the diagonal matrix of the eigenenergies of  
${\cal H}_{\rm LDA}$ with $\Bk$.
The matrix elements of $\Sigma(z)$ are given by calculating the self-energy operator term of eq.~(\ref{eq.DMFT-Hamiltonian}) 
based on eq.~(\ref{eq.psi}).

The DOS on the atomic $4f$ state is given by 
\Beqa
 \rho^{({\rm band})}_{\Gamma}(\VEP;\Bk)=-\frac{1}{\pi}\Im{\rm tr}
[\hat{O}_{\Gamma}G(\VEP+{\rm i}\delta;\Bk)],
\label{eq.rho-band}
\Eeqa
where the projection operator is defined as
\Beqa 
\hat{O}_{\Gamma}=\sum_{\Bi\gamma}
|\phi^{\rm a}_{\Bi(\Gamma\gamma)}><\phi^{\rm a}_{\Bi(\Gamma\gamma)}|.
\label{eq.density-op}
\Eeqa
The local DOS in the DMFT band calculation is obtained by summing $\rho^{({\rm band})}_{\Gamma}(\VEP;\Bk)$ over 
$\Bk$ in the Brillouin zone:
$\rho^{({\rm band})}_{\Gamma}(\VEP)=\frac{1}{N}\sum_{\Bk}\rho^{({\rm band})}_{\Gamma}(\VEP;\Bk)$~\cite{A41}. 
Here, $N$ is the total number of unit cells. 

The auxiliary impurity problem is solved by the NCA$f^{2}$vc method.
The splitting of the self-energy due to the SOI and CFS effects is considered. 
As shown in ref. \citen{A7}, 
this method  gives an accurate order of the Kondo temperature 
when the result is compared with that of the more correct numerical renormalization group (NRG) calculation~\cite{A7,A43} in a simple model case.

Since the method of the self-consistent calculation in the DMFT has been described previously~\cite{A6,A7}, 
we exclude the detailed explanation from this paper.
First of all, we calculate the self-consistent LDA band by the LMTO method, and 
potential parameters, except for the $f$ levels, 
are fixed to those in the LDA calculation.
(I) We calculate the atomic $4f$ density of states $\rho^{({\rm imp.})}_{\Gamma}(\VEP)$ ($4f$~DOS) for the auxiliary impurity 
Anderson model by the NCA$f^{2}$vc method with a trial energy dependence of the hybridization intensity (HI) 
and $4f$ levels\cite{D2},  
then calculate the local self-energy.
(II) The DMFT band calculation is carried out using the self-energy term, 
and the local $4f$~DOS in the DMFT band is calculated.
The calculation is iterated so that the $4f$~DOS of the local auxiliary 
impurity model and the DMFT band satisfy the self-consistent conditions~\cite{D1,A44}. 

The $4f$ level 
is adjusted 
in the DMFT self-consistent iterations under the condition that the  $4f$ occupation number has 
a given target value, $n_{f}({\rm rsl.target})$, 
which is estimated from the LDA band calculation.
The temperature dependence of the Fermi energy, $E_{\rm F}$, is neglected by fixing it at a value determined at a low temperature.
It is estimated using the occupation number of the renormalized band (RNB) calculation, in which the self-energy 
is approximated by an expansion form up to the linear term in the energy variable at $E_{\rm F}$ 
(see Appendix A for the calculation of the total occupation number).
The target $4f$ electron number $n_{f}({\rm rsl.target})$ is imposed on the occupation number
calculated directly using the resolvents to stabilize the self-consistency iterations. 
The occupation number obtained by the integration of the $4f$~DOS, $n_{f}({\rm intg.})$, has a deviation 
within 1.0\% 
from 
$n_{f}({\rm rsl.target})$  
because many intermediate calculation process 
are included.

\section{CePd$_{3}$
}

\subsection{Density of states}

%Fig1

\begin{figure}[htb]
\begin{center}
\includegraphics[width=8cm]{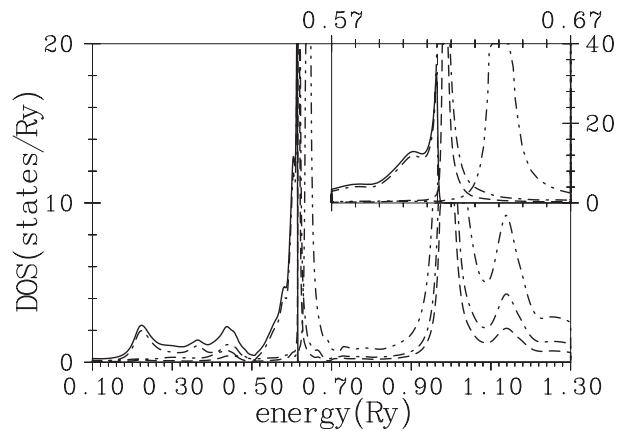}
\end{center}
\caption{
$4f$~DOS of CePd$_{3}$ at $T=37.5$~K.
The solid line shows the total $4f$ PES. 
The dashed line is the DOS of the ($j=5/2$)$\Gamma_{7}$ component, 
the dot-dash line is the DOS of the ($j=5/2$)$\Gamma_{8}$ component, and 
the two-dots-dash line is the DOS of the $j=7/2$ component.
The Fermi energy $E_{\rm F}=0.6143$~Ry is indicated by the vertical dot-dash line.
The inset shows spectra in the vicinity of $E_{\rm F}$.
}
\label{fig:CePd3-38-flsp}
\end{figure}

%Fig2

\begin{figure}[htb]
\begin{center}
\includegraphics[width=8cm]{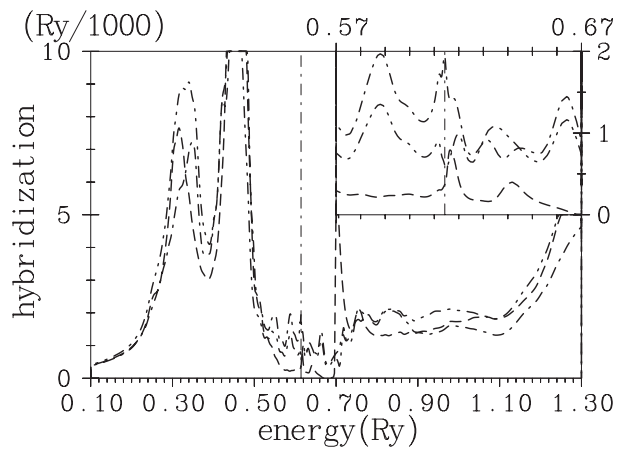}
\end{center}
\caption{
Hybridization intensity (HI)~\cite{D3} of CePd$_{3}$ at $T=37.5$~K.
The dashed line is the HI of the ($j=5/2$)$\Gamma_{7}$ component, 
the dot-dash line is the HI of the ($j=5/2$)$\Gamma_{8}$ component, and 
the two-dots-dash line is the HI of the $j=7/2$ component.
$E_{\rm F}$ is indicated by the vertical dot-dash line.
The inset shows HI in the vicinity of $E_{\rm F}$.
}
\label{fig:CePd3-38-mix}
\end{figure}

%Fig3

\begin{figure}[htb]
\begin{center}
\includegraphics[width=8cm]{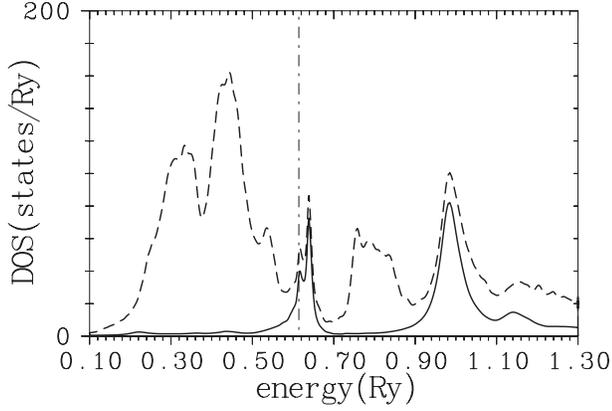}
\end{center}
\caption{
Total DOS (dashed line) and $4f$~DOS (solid line) of CePd$_{3}$ at $T=37.5$~K.
The spectra are broadened by an imaginary factor, $\gamma=0.01$~Ry, in the energy variable.
$E_{\rm F}$ is indicated by the vertical dot-dash line.
}
\label{fig:CePd3-38-bdos}
\end{figure}

%Table CePd3(813-1)

\begin{table}[t]
\caption{
Various quantities obtained in the DMFT calculation for CePd$_{3}$ at $T=18.75$~K.
$n_{\Gamma}^{({\rm imp.})}$ is the occupation number
in the auxiliary 
impurity problem of the effective HI of the DMFT calculation, 
$\VEP_{\Gamma}$ is the energy level in Ry.
$\rho_{\Gamma}(E_{\rm F})$ is the partial DOS at $E_{\rm F}$ in Ry$^{-1}$.
$\bar{Z}_{\Gamma}^{-1}=
 1-\partial\Re\Sigma_{\Gamma}(\varepsilon)/\partial\varepsilon|_{E_{\rm F}}$
is the mass renormalization factor of the $4f$ band,
$\bar{\VEP}_{\Gamma}$ is the effective energy of the renormalized band in Ry,
and $\bar{\Gamma}_{\Gamma}$ is the imaginary part of the self-energy in Ry.
The effective energy levels are measured from the Fermi energy $E_{\rm F}=0.6143$~Ry. 
The lattice constant is $a=7.80079$ a.u., and the spin-orbit interaction constant is  
$\zeta_{4f}=7.2302 \times 10^{-3} $ Ry.
The $4f$ level in the band calculation is $\VEP^{\rm LDA}_{4f}=0.6754$ Ry.
The electrostatic CFS of ($j=5/2$)$\Gamma_{7}$ is 
set to be $\Delta E_{\Gamma 7}(=\VEP^{\rm a}_{\Gamma 7}-\VEP^{\rm a}_{\Gamma 8})=150$~K. 
The target $4f$ occupation number is $n_{f}({\rm rsl.target})=0.98$, and   
the resultant occupation number calculated using the resolvent is 0.980.
The $4f$ occupation number calculated by integrating the spectra is $n_{f}({\rm intg.})=0.989$, and 
the obtained total band electron number is $N({\rm total} ; {\rm RNB})=34.006$. 
$E_{\rm inel}$ is the characteristic energy of the quasi-elastic excitation  
and $E_{\rm CFS}$ is the CFS excitation energy estimated from the magnetic excitation spectrum.
The occupation number of the $4f$ state in CePd$_{3}$ in LDA is $n_{f}({\rm Ce, LDA})=1.045$.
The Coulomb constant $U$ is set to be 0.51~Ry (6.9~eV)\cite{D2}.
}
\label{tab:CePd3}
\begin{halftabular}{@{\hspace{\tabcolsep}\extracolsep{\fill}}cccc} \hline
%\multicolumn{4}{c} {$r=1$} &\multicolumn{3}{c} {$ r=0.7$} \\
&$\Gamma_{7}$ & $\Gamma_{8}$ & $j=7/2$ 
\\ \hline
$n_{\Gamma}^{({\rm imp.})}$ & 
 0.084  & 0.719   & 0.185
                              \\
$\VEP_{\Gamma}$(Ry)                &
-0.19027  &-0.19182   &-0.16493
                               \\ 
$\rho_{\Gamma}(E_{\rm F})(Ry^{-1})$  &
3.8  &  59.0  & 0.5
                               \\ 
$\bar{Z}^{-1}_{\Gamma}$        & 
6.3     & 6.7 & 3.6 
                               \\
$\bar{\VEP}_{\Gamma}$(Ry)            &
0.0153  &0.0057 & 0.1184
                               \\  
$\bar{\Gamma}_{\Gamma}$(Ry)            &
2.7 $\times 10^{-4}$  & 16.0 $\times 10^{-4}$ & 3.0 $\times 10^{-4}$
                               \\ 
\multicolumn{3}{l}
 {$E_{\rm inel}=27$ meV, \hspace{0.5cm} $E_{\rm CFS}=41$ meV}  & \\
\hline
\end{halftabular}
\end{table}

In Fig. \ref{fig:CePd3-38-flsp}, we show the DOS of the $4f$ component ($4f$~DOS)
for CePd$_{3}$ at $T=37.5$ K. 
The solid line shows the total $4f$ component of the PES ($4f$~PES).
The dashed line is the DOS of the $\Gamma_{7}$ component and the dot-dash line is the 
DOS of the $\Gamma_{8}$ component.
The two-dots-dash line is the DOS of the $j=7/2$ component.
The CFS of the self-energy in the excited $j=7/2$ multiplet is neglected.
The vertical dot-dash line indicates the Fermi energy $E_{\rm F}=0.6143$~Ry.
The inset shows spectra in the energy region near $E_{\rm F}$.
The $4f$~DOS has a large peak at 0.642~Ry,  about 0.028~Ry~(0.38~eV) above $E_{\rm F}$.
This has mainly the $j=7/2$ character.
The spin-orbit splitting on the IPES side is usually enhanced in Ce systems.
The spectral intensity on the PES side  consists mostly of the $\Gamma_{8}$ component.
The PES has a sharp peak at $E_{\rm F}$ with a steep tail up to 0.094~Ry~(1.3~eV) below $E_{\rm F}$.
It also has a long tail with small structures reflecting the DOS of $4d$ states of Pd 
on the high-binding-energy side.
In the steep tail region, shoulders appear at binding energies of 0.004~Ry~(0.05~eV) and 0.024~Ry~(0.33~eV).
These may correspond, respectively, to the CFS and the SOI side band.
The sharp peak at $E_{\rm F}$ with the steep and the long tail has been observed 
in high-resolution experiments by Kasai {\it et al.}~\cite{A28}.
The shoulder due to the SOI seems to be identified.

The Kondo temperature $T_{\rm K}$ and the CFS excitation energy are, respectively, estimated to be about 27~meV~(310~K)
and 41~meV~(480~K) from magnetic excitation spectra, as 
will be shown in Fig. \ref{fig:CePd3-38-mag} in the next subsection.
We note that  spectra do not show appreciable change even when the temperature is decreased 
to 18.75~K in the calculation.

The parameters and the calculated values are given in Table \ref{tab:CePd3}.
The LMTO band parameters~\cite{A12} for states except for the $f$ component are fixed to those of the LDA calculation.
$E_{\rm F}$ is fixed to the value determined by the occupied state in the RNB, 
as is discussed in a later section.
The relative occupation number of the $\Gamma_{7}$ component to the $\Gamma_{8}$, 
0.084/0.719~(0.12) is  
small compared with the 0.5 expected from the ratio of the degeneracy, 
but is very large compared with the value expected from the simple model of the CFS for an isolated ion with 
$\VEP_{\Gamma 7}-\VEP_{\Gamma 8}=0.00155$~Ry~(240~K).
The ratio does not change so greatly even when the temperature is raised: 0.089/0.723~(0.12) at 150~K 
and 0.108/0.707~(0.15) at 300~K.
Moreover, the occupation of the $j=7/2$ component, 0.185, is not small, and is almost independent of $T$.
A simple picture of the CFS for an isolated ion cannot be applied.

Usually, the electrostatic potential causes cubic CFS in $4f$ electron systems 
with a higher energy level of about 150 K for the $\Gamma_{7}$ state~\cite{A47,A48}.
This is 
included in the present calculation. 
Even when it is neglected, 
the $4f$~DOS and the magnetic excitation spectrum are not greatly changed 
because the hybridization effect causes large CFS, of greater than 300~K~\cite{A49}. 

The target value of the occupation number on the atomic $4f$ states, $n_{f}({\rm rsl.target})$, has been tentatively chosen to be 
0.98 in the present calculation.
This value is small compared with the $4f$ occupation number, 1.045, of the LDA band calculation for CePd$_{3}$.
When we carry out the LDA calculation for compounds in which Ce ions are replaced by La ions, the occupation number on 
the $4f$(La) state usually amounts to 0.1~\cite{A50}.
We use the occupation number as reduced to 94\% of the LDA value as the atomic occupation number on the Ce $4f$ state.
Kanai {\it et al.} concluded that $4f$ occupancy in CePd$_{3}$ is expected to be 0.92 on the basis of the results of 
resonant inverse photoemission (RIPES) experiments~\cite{A25,A45,A46}.
When we perform the DMFT band calculation by setting $n_{f}({\rm rsl. target})$ to be 0.92, $T_{\rm K}$ 
determined from the magnetic excitation spectrum is expected to be 500~K. 
On the other hand, $T_{\rm K}$ is greatly reduced to about 10~K when we choose 1.05.

In Fig. \ref{fig:CePd3-38-mix}, we show the trial HI~\cite{D3} obtained in the DMFT band calculation at 37.5 K.
It has very large peaks corresponding to the $4d$ band of Pd, 
but these peaks are located in the energy region deep below $E_{\rm F}$.
The HI is not high in the energy region near $E_{\rm F}$.
In particular, the HI of the $\Gamma_{7}$ component is low though it is high in the deeper energy region.
The overall features of the HI in the DMFT band are similar to those calculated directly using the LDA band,
but the HI in DMFT is increased in the vicinity of $E_{\rm F}$ to about twice the LDA value for 
the $\Gamma_{7}$ and $\Gamma_{8}$ components, and is decreased in the deep energy region.
Moreover, 
the DMFT calculation causes fine structures of the HI in the vicinity of $E_{\rm F}$, which are shown in the inset of the 
figure.
This contrasts with the 
HI in the LDA, which has a weak energy dependence in this region.
The HI of the $\Gamma_{8}$ component in DMFT has a small peak at $E_{\rm F}$, while those of the $\Gamma_{7}$ and $j=7/2$ components have 
small peaks above or on both sides of $E_{\rm F}$.
The reason for this different behavior is not clear at present, but 
we note that the peaks of the $4f$~DOS for the latter two cases are located above $E_{\rm F}$.
When we do a calculation in the fictitious case that $\Gamma_{7}$ is mainly occupied by assigning a low energy level to it, 
the HI of $\Gamma_{7}$ has a small peak at $E_{\rm F}$.
However, this result should not be used as a general rule, because the modification of HI in DMFT is delicately dependent on 
details of the band structures.
The Kondo temperature is increased in the DMFT calculation in the CePd$_{3}$ case.
We have obtained the Kondo temperature of 10 K in the single impurity calculation using 
the HI of the LDA band.

The total DOS at $T=37.5$ K is shown by the dashed line in Fig. \ref{fig:CePd3-38-bdos}.
The large peaks at about 0.3 and 0.45 Ry have the $4d$ character of Pd, and that at 0.8 Ry has the $5d$ character of Ce.
These peaks are also obtained by the LDA calculation~\cite{A51}.
The sharp $4f$ peaks slightly above $E_{\rm F}$, which are called the $f^{1}$ peak in IPES, 
are also obtained in the LDA calculation\cite{A40}.
Their 
intensity in DMFT is reduced compared with that in the LDA  because a part of it is transferred to the intensity of the peak 
at about 1.0 Ry, which is called the $f^{2}$ peak. 
In the present calculation, the width of the $f^{2}$ peak is not large because the multiplet splitting of the $f^{2}$ final 
state is neglected.

In the analysis of RIPES experiments, the ratio of the $f^{1}$ peak to the total RIPES intensity has been given as
0.22~\cite{A25}, 
whereas it is estimated to be about 0.2 
in the present calculation. 
The present DMFT calculation seems to give results that emphasizes the hybridization effects 
strongly (i. e. the higher $T_{\rm K}$).

\subsection{Magnetic excitation}

%Fig4

\begin{figure}[htb]
\begin{center}
\includegraphics[width=8cm]{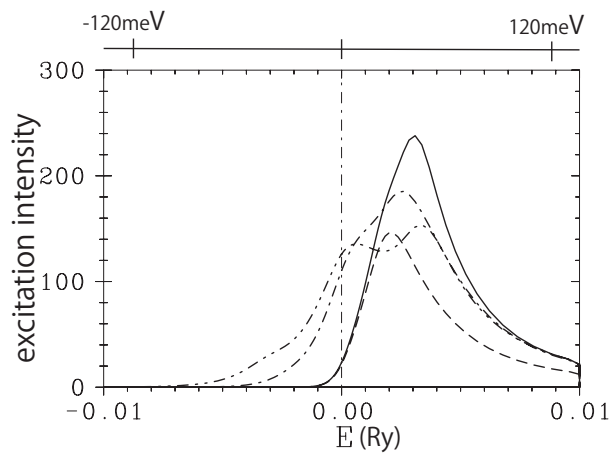}
\end{center}
\caption{
Magnetic excitation spectrum of CePd$_{3}$. 
The solid line shows the spectrum at $T=37.5$ K, 
the dot-dash line is the spectrum at $T=150$ K, and 
the two-dots-dash line is the spectrum at $T=300$ K.  
The dashed line is the spectrum in a fictitious case where matrix elements of the magnetic moment are restricted 
within the intra-$\Gamma_{8}$ manifold of space.
}
\label{fig:CePd3-38-mag}
\end{figure}

In Fig. \ref{fig:CePd3-38-mag} we show the magnetic excitation spectrum.
The total magnetic excitation spectrum at $T=37.5$ K is shown by the solid line.
It has a peak at about $E=0.003$ Ry (41 meV).
The dashed line depicts the spectrum for a fictitious case in which the matrix elements of the magnetic moment are nonzero only in the 
manifold of $\Gamma_{8}$, and thus it may correspond to the excitation spectrum within the $\Gamma_{8}$ manifold.
It has a peak at about $E=0.002$ Ry (27 meV), 
and leads to a faint shoulder in the solid line.
The CFS excitation energy seems to be 
slightly larger than $T_{\rm K}$ in the present calculation.
The calculated magnetic susceptibility is 1.9~$\times 10^{-3}$~emu/mol, whereas the experimental value is 
1.5~$\times 10^{-3}$~emu/mol at $T \sim 0$~K~\cite{A52,A56}.
We show magnetic excitation spectra at $T=150$~K and at $T=300$~K by the dot-dash line and the two-dots-dash line, respectively.
The spectrum at $T=300$~K has a broad peak centered at about 0.003~Ry~(41 meV). 
The shoulder shifts to $E \sim 0$ and becomes a peak. 
the overall features do not change so greatly when we neglect the electrostatic CFS of 150~K.
The magnetic excitation spectrum observed in the wave-vector-integrated case has a peak at about 55~meV at $T= 10$~K, 
and the peak shifts to the 
low-energy side as T increases~\cite{A19}.
The calculated results seem to be generally consistent with those of the experiment. 
However, the peak at $E \sim 0$ is higher at $T=300$~K in the experiments~\cite{A19}.
The present DMFT calculation seems to give stronger HI compared with the value in the experiments.
The detailed calculation of physical quantities using finely tuned parameters will be given in the future.

In ref.~\citen{A20}, excitation peaks with an energy of 15~meV and less than~3 meV are indicated. 
Low-energy peaks are not expected in the present calculation of the wave-vector-integrated spectra because $T_{\rm K}$ is not low.
One possibility of the origin of the low-energy peaks may be the wave vector dependence of the magnetic 
excitation spectra, as noted in ref.~\citen{A21}.

\subsection{RNB, and wave number-vector-dependent DOS}

%Fig5

\begin{figure}[htb]
\begin{center}
\includegraphics[width=10cm]{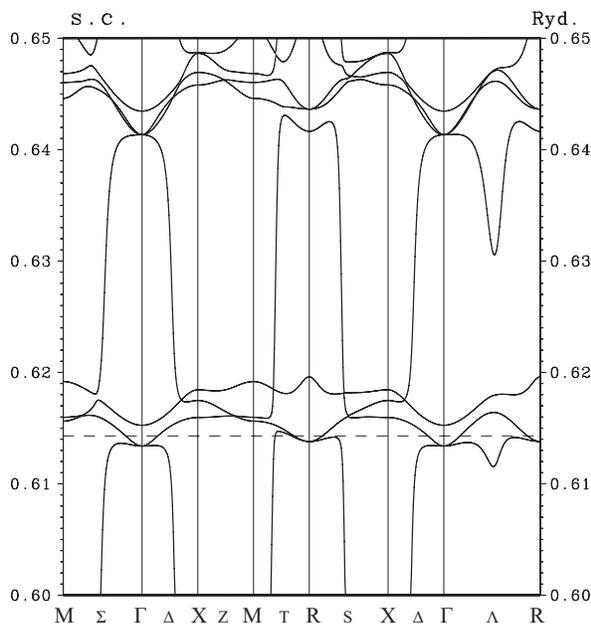}
\end{center}
\caption{
Band dispersion of the renormalized band (RNB) picture for CePd$_{3}$ at $T=37.5$ K. 
The symbols under the horizontal axis denote the symmetry points and axes of the Brillouin zone of the $sc$ lattice.
$E_{\rm F}$ is indicated by the horizontal dashed line.
}
\label{fig:CePd3-rnb}
\end{figure}

%Fig6

\begin{figure}[htb]
\begin{center}
\includegraphics[width=8cm]{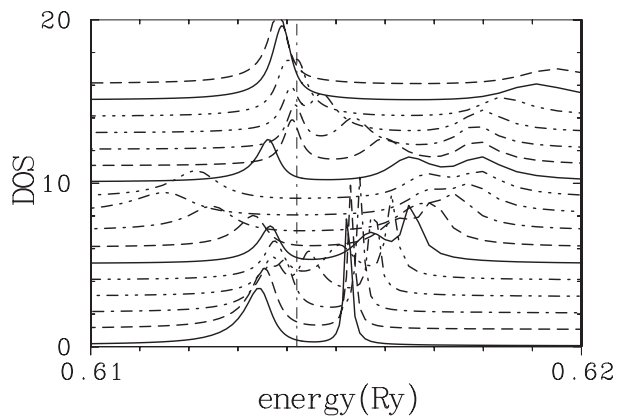}
\end{center}
\caption{
Wave number vector ($\Bk$) dependence of the DOS (k-DOS) for CePd$_{3}$ at $T=37.5$ K.
$\Bk$ is moved from the $\Gamma$ point (bottom) to the R point (top)  along the $\Lambda$ axis. 
The total DOS is shown, but the $4f$ component is dominant in this energy region of CePd$_{3}$.
}
\label{fig:CePd3-38-kspc}
\end{figure}

%Fig7

\begin{figure}[htb]
\begin{center}
\includegraphics[width=8cm]{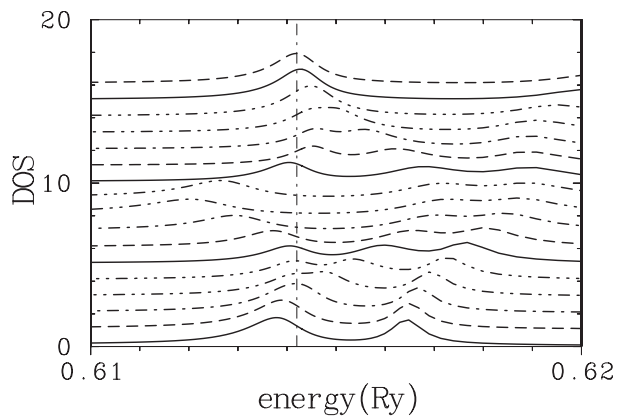}
\end{center}
\caption{
k-DOS of CePd$_{3}$ at $T=150$ K.
$\Bk$ is changed from the $\Gamma$ point (bottom) to the R point (top)  along the $\Lambda$ axis.
}
\label{fig:CePd3-300-kspc}
\end{figure}

%Fig8

\begin{figure}[htb]
\begin{center}
\includegraphics[width=8cm]{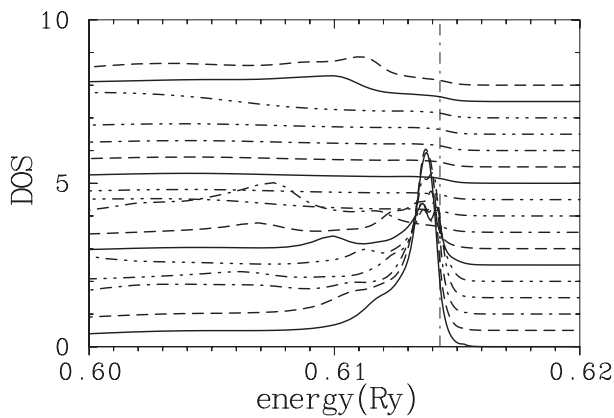}
\end{center}
\caption{
Angle-resolved PES of CePd$_{3}$ at $T=37.5$ K.
The surface is assumed to be (111). 
The representative $\Bk$ is swept from the $\Gamma$ (bottom) point to the M (top) point along the $\Sigma$ axis, 
and the intensities are averaged over the  
wave number vector normal to the (111) surface.
This means that $\Bk$ also runs from the R (bottom) point to the X (top) point along the S axis. 
The Fermi energy $E_{\rm F}$ is indicated by the vertical dot-dash line.
The total DOS is shown, but the intensity above the energy of 0.612 Ry is dominated by the $4f$ character. 
The intensity below 0.611 Ry has mainly the non-$4f$ character. 
}
\label{fig:CePd3-38-PES}
\end{figure}

%Fig9

\begin{figure}[htb]
\begin{center}
\includegraphics[width=8cm]{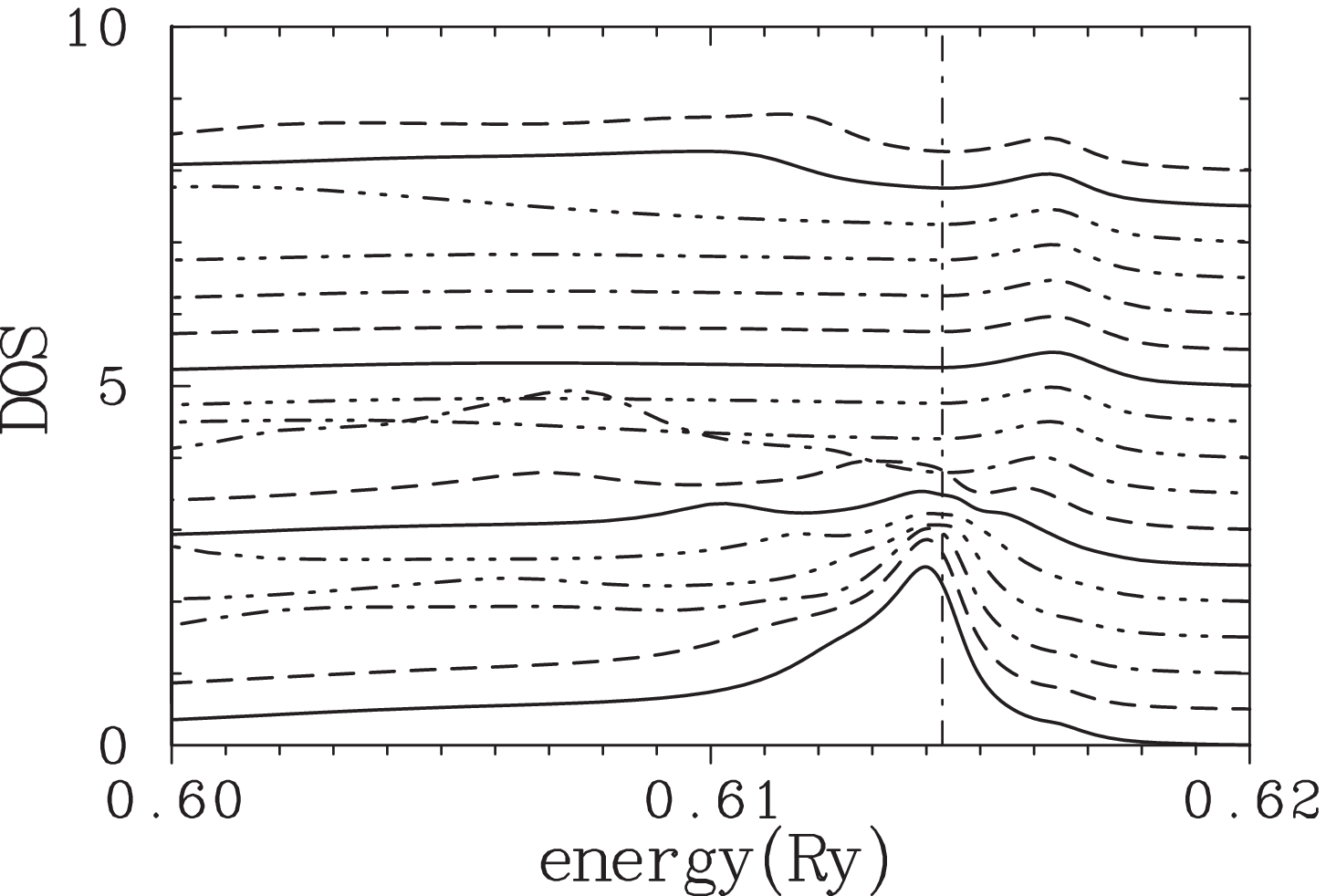}
\end{center}
\caption{
Angle-resolved PES of CePd$_{3}$ at $T=150$ K.
For the definition of lines, see the caption of Fig. \ref{fig:CePd3-38-PES}. 
}
\label{fig:CePd3-300-PES}
\end{figure}

In Fig. \ref{fig:CePd3-rnb}, we show the RNB dispersion at $T=37.5$ K.
The energy shift (the real part of the self-energy at $E_{\rm F}$: $\Re\Sigma_{\Gamma}(E_{\rm F}$)) and the mass renormalization factor
($1-\partial\Re\Sigma_{\Gamma}(\varepsilon)/\partial\varepsilon|_{E_{\rm F}}$), which are 
given in Table \ref{tab:CePd3}, are 
taken into account in this calculation.
Narrow bands with the $j=5/2$ character appear slightly above $E_{\rm F}$, and 
those with the character of $j=7/2$ appear around the energy 0.645 Ry which is 
near the energy of the $j=7/2$ peak in the $4f$~DOS shown in Fig. \ref{fig:CePd3-38-flsp}.

The lowest $4f$ band sinks to below $E_{\rm F}$ near the $\Gamma$ and R points,
and is located above $E_{\rm F}$ in the other regions.
The dispersion of RNB corresponds well with the behavior of the $\Bk$-dependent density of states (k-DOS).
For example, we show the k-DOS when $\Bk$ moves from the $\Gamma$ (bottom) to the R (top) point along the $\Lambda$ line 
in Fig. \ref{fig:CePd3-38-kspc}. 
A peak of the DOS with mainly the $4f$ character is located below $E_{\rm F}$ at the $\Gamma$ and R points.
Starting from the peak below $E_{\rm F}$ at the $\Gamma$ point, one of ridge lines runs above the Fermi energy 
across $E_{\rm F}$, and then connects to the peak below $E_{\rm F}$ at the R point.
Another runs to the low-energy side of $E_{\rm F}$ up to 0.611 Ry at the halfway, and then turns back to connect to the peak 
below $E_{\rm F}$ at the R point.
This ``hanging'' branch does not cross the Fermi energy.

Note that we have depicted the total spectral intensity, not the $f$-component,  in Fig. \ref{fig:CePd3-38-kspc}.
In the energy region shown in the figure, the spectral intensity has mainly the $4f$ character. 
On the other hand, very sharp spectral peaks of a non-$4f$ character appears in the energy region outside of the 
figure.
The hanging band on the $\Lambda$ axis has a stronger non-$4f$ character.  
This branch is a hybridization band between the $4f$ of Ce and a 
conduction band that has the character of the $sp$-free electron band and the $5d$ of Ce.

The dispersion of the RNB is qualitatively similar to the result of the band structure determined by Hasegawa and Yanase by the 
LDA calculation~\cite{A53},
although the width of the $4f$ band with $j=5/2$ is about 0.005~Ry in the RNB, whereas that of the LDA band is about 0.02~Ry.
Both calculations give electron pockets at the $\Gamma$ and R points, and hole pockets centered on the T axis~\cite{D4}.

The Main features of the band dispersion near the Fermi energy are formed by the hybridization of the narrow $4f$ bands and the wide 
$sp$ bands.
The $4f$ bands have dispersion characterized by the LMTO (linear combination of atomic orbitals) tight-binding bands of the $sc$ lattice.  
Although the $4f$ band width is reduced in DMFT, 
the qualitative features of the dispersion of the $4f$-$sp$ hybridized bands are not changed because the number of 
participating $sp$ bands is small and their dispersion is very rapid compared with that of $4f$ bands.

%Main features of the band dispersion near the Fermi energy are formed by the hybridization of the narrow $4f$ bands and the wide 
%$sp$ bands.
%The $4f$ bands have dispersion characterized by the LCAO tight-binding bands of the $sc$ lattice , 
%and are located with nearly one electron occupancy at the Fermi energy of the $sp$ bands.
%The $4f$ band width is reduced in the DMFT.
%However, qualitative features of the dispersion of the $4f$-$sp$ hybridized bands are not changed because the number of 
%participating $sp$ bands is small and the dispersion of them is very steep compared with that of $4f$ bands.

Here, we must note a weak point of the NCA$f^{2}$vc method, that is it does not automatically ensure  the Fermi liquid sum rule,
 i.e., the 
integral of the total DOS below $E_{\rm F}$, 
$N({\rm total}; \rho)=\int{\rm d}\VEP\frac{1}{N}\sum_{\Bk}(-\frac{1}{\pi}\Im G(\VEP+{\rm i}\delta;\Bk))f(\VEP)$, 
is not equal to the occupation number of electrons
calculated by the volume of the occupied states in the $\Bk$ space. 
The quantity $N({\rm total}; \rho)$ is a smaller value in CePd$_{3}$, and thus we 
obtain a higher Fermi energy.
If we use it in the RNB calculation, the volume of the occupied states in the $\Bk$ space becomes large.
In the case of CePd$_{3}$, which has an even total electron number,  
the balance between the electron and hole states is lost.
In this study, we tentatively use the Fermi energy determined using the quantity
$N({\rm total};{\rm RNB})=\frac{1}{N}\sum_{\lambda\Bk}f(E_{\rm RNB}(\lambda\Bk))$ 
to ensure the electron-hole balance in the RNB band, where $E_{\rm RNB}(\lambda\Bk)$ is 
the energy obtained by the RNB calculation. 
(For more details, see Appendix B.)

When we calculate the RNB dispersion at $T=150$ K, the $4f$ band shifts slightly to the high-energy side.
The $4f$ state at the $\Gamma$ point nears $E_{\rm F}$, and the $4f$ state at the R point is located on $E_{\rm F}$. 
Therefore, a part of the hanging band on the $\Lambda$ line rises above $E_{\rm F}$. 
Hole pockets on the T axis grow into larger hole regions around the R point.  
As the temperature increases further, the $4f$ state at the R point shifts up to above $E_{\rm F}$, and a large hole surface 
enclosing the R point appears.
In other words, we have a connected electron Fermi surface (FS) that contains the X and M points inside it.

The primary structure of the FS of the RNB at high temperatures is similar to that of the FS of 
the LDA band of LaPd$_{3}$, 
but their fine topologies will differ.
At $T=150$~K, the main part of the hanging band on the $\Lambda$ line 
is still located below $E_{\rm F}$. 
The hole sheets around the $\Gamma$ and R points are separated by an electron region on the $\Lambda$ axis.
On the other hand, 
this entire hanging branch is located 
above $E_{\rm F}$ in LaPd$_{3}$, and thus the separation by the electron
region does not occur.

We should, of course,  note that the RNB picture has only limited meaning at high temperatures because the imaginary part 
is large~\cite{D5}. 
The k-DOS at $T=150$~K for $\Bk$ along the $\Lambda$ line is shown in Fig. \ref{fig:CePd3-300-kspc}. 
The widths of peaks become large, 
but we can see that the peak at the R point is located almost at $E_{\rm F}$, 
and the peak at $\Gamma$ nears $E_{\rm F}$. 
The trace of broad peaks shows a shift corresponding to that of the RNB dispersion. 
We can recognize that the ridge of the peaks of the hanging band crosses the Fermi energy, as noted in the 
previous paragraph.
At $T=300$~K, the width of peaks becomes so large that it surpasses the fine $\Bk$ dependence of the spectra.
However, 
the $4f$ peak near the Fermi energy still exists in the $4f$~DOS calculation, similar to that 
shown in Figs. \ref{fig:CePd3-38-flsp} and \ref{fig:CePd3-38-bdos}.
It moves slightly to the high-energy side with increasing  width at 300 K.

The present calculation has been performed with fixed $E_{\rm F}$ and $n_{f}({\rm rsl.target})$.
The calculated total occupation number in the DMFT band increases to $N({\rm total} ; {\rm RNB})=34.304$ at $T=300$~K,   
although the energies of the $4f$ bands shift upward~\cite{D6}.  
We point out that
even when we move $E_{\rm F}$,  
the energy of $4f$ bands relative to the Fermi energy will not change greatly 
because the Kondo resonance peak usually shifts following the change of the Fermi energy.   

Recently, the wave vector dependence of the PES has been observed under the $4d \rightarrow 4f$ resonant condition~\cite{A29}.
The experiment was carried out for the (111) surface of a thin film by sweeping the $\Bk$ vector along 
$\bar{\Gamma}$-\={K}-\={M}-\={K}-$\bar{\Gamma}$ in the surface Brillouin zone (BZ). 
This may correspond to the following sweeping of $\Bk$ in the 3-dimensional {\it sc} BZ: the component parallel to the surface 
runs as  $\Gamma$-($\Sigma$)-M-($\Sigma$)-$\Gamma$ with averaging over its components normal to the surface.
This means that the sweeping of the parallel component  R-(S)-X-(S)-R is also included.

In Fig. \ref{fig:CePd3-38-PES}, we show the PES when the representative wave vector moves from the $\Gamma$ (bottom) 
point to the M (top) point
with average intensities for $\Bk$ normal to the surface.
The total intensity is plotted in the figure, but the intensity above the energy of 0.612 Ry has mainly the $4f$ character.
The spectra have a peak for $\Bk$ near $\Gamma$, and this corresponds well to the experimental results.
The intensity below the energy of 0.611 Ry has a non$4f$ character.
As noted previously, the contribution from the R point is also included at the representative $\Gamma$ point.
At these points, the $4f$ component is located below $E_{\rm F}$ at $T=37.5$ K.  
In Fig. \ref{fig:CePd3-300-PES}, we show spectra at $T=150$ K.
The peak of the intensity at $\Gamma$ nears $E_{\rm F}$ because of the shifting up of the $4f$ band. 
The thermal distribution effect and the very high intensity of the $4f$~DOS above $E_{\rm F}$ also have affect  
this spectral shape.
The change of the peak can be checked in experiments.

\section{CeRh$_{3}$
}
%\input{section3.tex}

%\subsection{density of states and the magnetic excitation}

%Fig10

\begin{figure}[htb]
\begin{center}
\includegraphics[width=8cm]{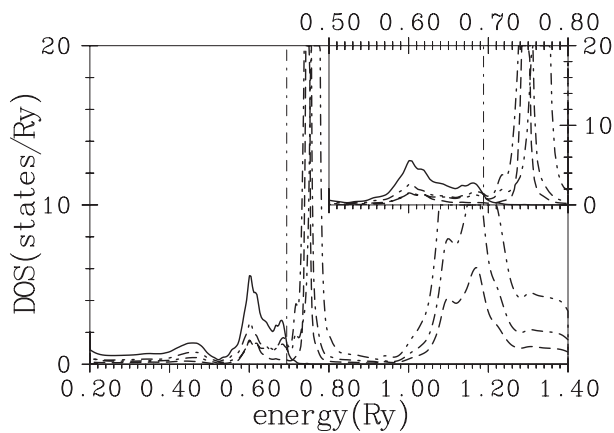}
\end{center}
\caption{
$4f$~DOS of CeRh$_{3}$ at $T=10^{3}$ K.
For the definition of lines, see the caption of Fig.~\ref{fig:CePd3-38-flsp}.
$E_{\rm F}=0.6940$~Ry is indicated by the vertical dot-dash line.
}
\label{fig:CeRh3-flsp}
\end{figure}

%Fig11

\begin{figure}[htb]
\begin{center}
\includegraphics[width=8cm]{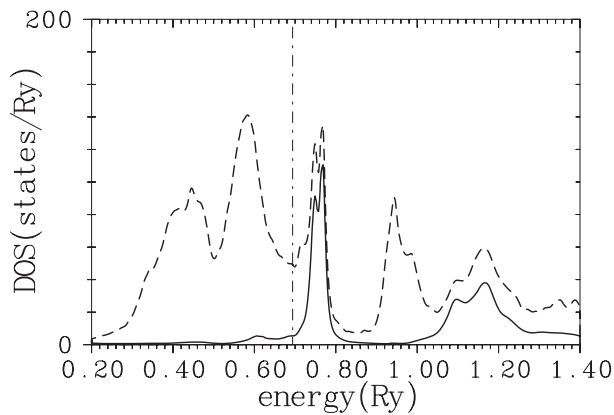}
\end{center}
\caption{
Total DOS (dashed line) and $4f$~DOS (solid line) of CeRh$_{3}$ at $T=10^{3}$ K.
See the caption of Fig.~\ref{fig:CePd3-38-bdos}.
}
\label{fig:CeRh3-bdos}
\end{figure}

%Fig12

\begin{figure}[htb]
\begin{center}
\includegraphics[width=8cm]{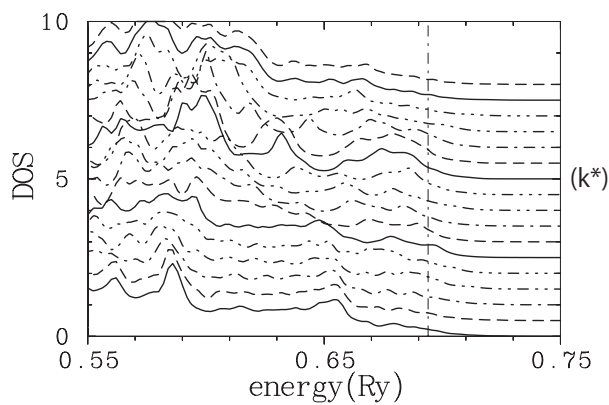}
\end{center}
\caption{
Angle-resolved PES of CeRh$_{3}$ at $T=300$ K.
For the definition of lines, see the caption of Fig.~\ref{fig:CePd3-38-PES}.
The $\Bk$ point corresponding to the second solid line from the top is named $\Bk^{*}$ in the text. 
}
\label{fig:CeRh3-kpes}
\end{figure}

%Fig13

\begin{figure}[htb]
\begin{center}
\includegraphics[width=8cm]{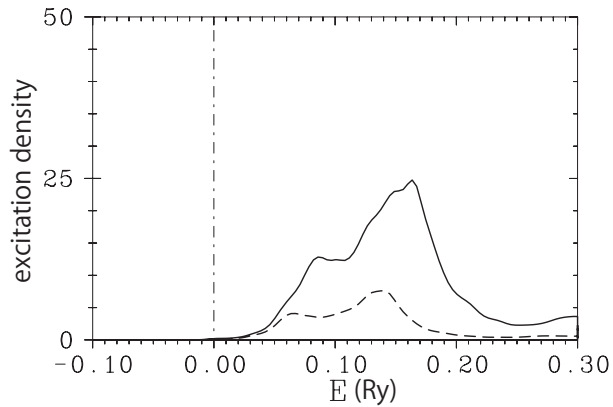}
\end{center}
\caption{
Magnetic excitation spectra of CeRh$_{3}$ at $T=10^{3}$ K.
The solid line is the total intensity and the dashed line is the result calculated using  a fictitious model in which 
the matrix elements of the magnetic moment are restricted within the $j=5/2$ manifold of states.   
}
\label{fig:CeRh3-mag}
\end{figure}

%Table CeRh3(211-1)

\begin{table}[t]
\caption{
Various quantities obtained in the DMFT calculation for CeRh$_{3}$ at $T=10^{3}$ K.
For the definition of notation, see the caption of Table \ref{tab:CePd3}.
The Fermi energy is $E_{\rm F}=0.6940$~Ry and $\Delta E_{7}=0$ K.
The lattice constant is $a=7.6005$ a.u., and the spin-orbit interaction constant is  
$\zeta_{4f}=7.3374 \times 10^{-3}$ Ry.
The $4f$ level in the band calculation is $\VEP^{\rm NCA}_{4f}=0.7639$ Ry.
The target $4f$ occupation number is $n_{f}({\rm rsl.target})=0.94$, 
and the resultant occupation number calculated using the resolvent is 0.940.  
The $4f$ electron number calculated by the integrating the spectra is
$n_{f}(\rm intg.)=0.943$, 
and the obtained total band electron number is $N({\rm total} ; {\rm RNB})=30.994$.
$n_{f}({\rm Ce,LDA})=0.996$.
}
\label{tab:CeRh3}
\begin{halftabular}{@{\hspace{\tabcolsep}\extracolsep{\fill}}cccc} \hline
%\multicolumn{4}{c} {$r=1$} &\multicolumn{3}{c} {$ r=0.7$} \\
&$\Gamma_{7}$ & $\Gamma_{8}$ & $j=7/2$ 
\\ \hline
$n_{\Gamma}^{({\rm imp.})}$ & 
 0.178  & 0.319   & 0.446
                              \\
$\VEP_{\Gamma}$(Ry)                &
-0.26923  &-0.26921   &-0.24426
                               \\ 
$\rho_{\Gamma}(E_{\rm F})$ (Ry$^{-1}$)                &
0.27  &  1.40 & 0.99
                               \\ 
 $\bar{Z}^{-1}_{\Gamma}$        & 
2.2    & 1.9 & 1.9 
                               \\
$\bar{\VEP}_{\Gamma}$(Ry)            &
0.1218  &0.1216 & 0.1646
                               \\  
$\bar{\Gamma}_{\Gamma}$(Ry)            &
 $4.6 \times 10^{-3}$  & $12.6 \times 10^{-3}$ & $5.8 \times 10^{-3}$
                               \\ 
\multicolumn{3}{l}
 {$E_{\rm inel}=0.03$ Ry }  & \\
\hline
\end{halftabular}
\end{table}

In Fig. \ref{fig:CeRh3-flsp}, we show the $4f$~DOS of CeRh$_{3}$ at $10^{3}$~K.
The $4f$ spectrum has a sharp and large peak at an energy about of 0.76 Ry, which is 0.066 Ry (0.9 eV) above the Fermi energy.
This separation of energy from $E_{\rm F}$ is slightly smaller than that of the $4f$ band in the LDA, about 0.009 Ry (1.2 eV). 
The result seems to be consistent with those of RIPES experiments and their detailed analysis~\cite{A24,A30}.
The PES has a relatively large peak at about 0.1 Ry (1.4 eV) below $E_{\rm F}$ and a peak at $E_{\rm F}$. 
The qualitative behavior of the present PES is similar to the result obtained by Harima in the LDA~\cite{A54}. 
However, the binding energy of the peak at 0.1~Ry below $E_{\rm F}$ is slightly lower and the intensity is 
higher than that of the LDA.
The peak at $E_{\rm F}$ is also slightly sharper.
In the experiment, the sharp peak at $E_{\rm F}$ was observed, but the peak at 0.1 Ry below $E_{\rm F}$ has not been 
identified at present~\cite{A31,A32,A33,A34}.

In Fig. \ref{fig:CeRh3-bdos}, we show the total and $4f$~DOS of CeRh$_{3}$ at $T=10^{3}$~K. 
The $4d$~band of Rh exhibits strong peaks of the DOS at about 0.4 and 0.6 Ry.
The $4f$ component also has  a small peak at about 0.6 Ry, as noted previously. 
The Fermi energy is located in the top region of the $4d$ band,  and the hybridization intensity in this region is high.
Dispersions of the RNB are almost identical to those of the LDA in the energy region very near $E_{\rm F}$,
but the width of the $4f$ band, which is located at 0.76~R, is about 2/3 that of the LDA.
A $f^{2}$ satellite peak appears in DMFT at about 1.2 Ry 
on the high-energy side.
The ratio of the intensity of peak at 0.76 Ry ($f^{1}$ peak) to the total IPES intensity is estimated to be about 0.4 in 
the present calculation, 
while a slightly larger value, 0.6, was obtained in the experiment. 
Uozumi {\it et al.} predicted the $4f$ occupation number to be 0.86~\cite{A30}, 
but we tentatively used $n_{f}({\rm rsl.target})=0.94$, which is 94\% of the LDA value.
The difference between these values is not small, but will not cause extreme differences in the 
calculation of the very strong hybridization limit. 
The mass enhancement factor is expected to be about 2, as given in Table \ref{tab:CeRh3}.

In Fig.~\ref{fig:CeRh3-kpes}, we show the PES for the (111) surface when the representative wave vector moves 
from the $\Gamma$ point (bottom) to the M point (top).
In the energy region between $E_{\rm F}$ and 0.6~Ry, the relative intensity of the $4f$ component is about 10\% 
of the total.
The spectra have fine peaks as if some flat bands exist slightly below $E_{\rm F}$.
However, we note that no flat bands exist very near $E_{\rm F}$ in the RNB dispersion.
Let us denote as $\Bk^{*}$ the representative $\Bk$ corresponding to the second solid line from the 
top~(i.e., the $\Bk$ point on the $\Sigma$ axis at a distance of about 0.3 $\times$ length of the $\Sigma$ axis
from the M point).
The spectral intensity slightly below $E_{\rm F}$ is relatively large for $\Bk$ near $\Bk^{*}$.
For the wave vectors near $\Bk^{*}$, several bands stay slightly below $E_{\rm F}$ around the M and X points 
when the normal components are varied.
Calculated results show quantitatively similar behaviors to experimental results, but the careful separation 
of the surface and bulk components is necessary to enable a detailed comparison~\cite{A55}.

The magnetic excitation spectra are shown in Fig. \ref{fig:CeRh3-mag}.
They have a steep increase at the excitation energy 
of about 0.03 Ry.
This energy may correspond to the energy from $E_{\rm F}$ to the low-energy edge of the peak at  0.76 Ry of the $4f$~DOS.
The-low energy end of the spectrum mainly originates from the excitation within the $j=5/2$ components, 
but contribution of the $j=7/2$ components is not small even in the low-excitation-energy region.
The $j=7/2$ components also join the Kondo effect. 
We may expect the Kondo temperature of this system to be about 0.03~Ry  (4700~K).
The calculated value of the magnetic susceptibility is $0.7 \times 10^{-4}$~emu/mol.
The experimental magnetic susceptibility ($\chi \sim 4 \times 10^{-4}$ emu/mol)~\cite{A56} may indicate a high $T_{\rm K}$ of 
several thousand K, 
but the calculated value of 4700~K seems to be too high.
DOSs, both of the total and of the $4f$ component, do not have any gap, as seen in Fig. \ref{fig:CeRh3-flsp}, 
but the magnetic excitation 
spectrum has a shape indicating the existence of a pseudogap. 
We note that, in Fig.~\ref{fig:CeRh3-mag}, the excitation spectra of only intra-$4f$ components are shown.
A broad continuous $5d$ component will be superposed on these spectra. 

The HI of CeRh$_{3}$ in the LDA calculation has a spectrum shape similar to that of the partial DOS on the 
$5d$ state of Rh, i.e., the 
spectral shape given by subtracting the $4f$ and $5d$ parts of the Ce ion from 
the total DOS in Fig.~\ref{fig:CeRh3-bdos}. 
The HI in DMFT is almost equal to that of the LDA in the high-energy region, 
but it has a steep dip at an energy slightly above $E_{\rm F}$.
Similar behavior has been seen in the HI of the $j=7/2$ component of CePd$_{3}$ which has shown 
in Fig.~\ref{fig:CePd3-38-mix}.

\section{Summary and Discussion}

We have studied the electronic structures of CePd$_{3}$ and CeRh$_{3}$ 
on the basis of the DMFT calculation.
The auxiliary impurity problem was solved by a method named NCA$f^{2}$vc, which includes the correct exchange process
of the $f^{1} \rightarrow f^{0}$ and $f^{1} \rightarrow f^{2}$ virtual excitation.
The splitting of the self-energy owing to the SOI and CFS effects was also considered.

The DMFT band calculation gives Fermi surface structures similar to those obtained by the LDA calculation in CePd$_{3}$~\cite{A53} 
at very low temperatures.
Electron pockets appear at the bottom of the $4f$ band at the $\Gamma$ and R points.
Hole pockets appear that are centered on the T symmetry axis.

The $4f$ band shifts to the high-energy side relative to the Fermi energy as the temperature increases. 
At the same time, the lifetime width of the $4f$ states increases.
The band structures produced by the band overlap between the $4f$ and non-$4f$ components shift up  
to the high energy side of the Fermi energy.
Therefore the primary structures of the band in the vicinity of the Fermi energy approaches to those of the $sp$ free-electron-like 
band of LaPd$_{3}$. 
However, some characteristic features of the LDA band of CePd$_{3}$ remain at higher temperatures. 
For example, a region that electrons occupy will appear on the $\Lambda$ axis in CePd$_{3}$ at $T \sim 150$~K, which 
is not expected in LaPd$_{3}$.
In CePd$_{3}$, the lifetime broadening overcomes the $\Bk$ dependence of the $4f$ spectrum at room temperature, 
thus, the $4f$ state becomes a broad dispersionless state located above $E_{\rm F}$.

The ARPES of CePd$_{3}$ shows strong intensity near the representative $\Gamma$ point at low temperatures 
because the $4f$ band is located below the 
Fermi energy at the $\Gamma$ and R points. 
This result seems to be consistent with the recent experimental results~\cite{A29}.
The observed $4f$ component will be greatly reduced at room temperature because of the shift of the $4f$ band 
to the high-energy side.

The k-integrated magnetic excitation spectrum has a peak at 41~meV and a faint shoulder structure at about 27~meV at low 
temperatures.
The temperature dependence of the excitation spectrum generally seems to be consistent with the results of experiments~\cite{A19}.
The magnetic CFS excitation energy is estimated to be about 41~meV, while the Kondo temperature is slightly smaller,  
320~K~(27~meV).

The calculated PES shows good correspondence with the bulk component obtained in recent high-resolution experiments~\cite{A28}.
The intensity ratio of the $f^{1}$ peak of the IPES to the total IPES is estimated to be about 0.2, while it was  
predicted to be 0.22 by the recent experiment analysis~\cite{A25}.
The HI is enhanced near the Fermi energy in the DMFT band compared with that of 
the LDA calculation.
The present DMFT calculation seems to give electronic structures with slightly stronger HI than that expected from 
experiments for CePd$_{3}$.

The DMFT calculation for  CeRh$_{3}$ gives an almost identical band structure to that obtained by the 
LDA calculation.
However, the energy of the $f^{1}$ peak of the IPES in the former is slightly lower than that in the latter.
In addition, the $f^{2}$ satellite peak appears in the DMFT band calculation.
The calculated intensity ratio of the $f^{1}$ peak to the total IPES, 0.4, is comparable to, but smaller than the 
experimental value of 0.6.~\cite{A30}. 
The calculated PES has a sharp peak at the Fermi energy. 
A peak reflecting the $4d$(Rh) DOS also appears, similarly to the result of the LDA calculation.
The former has been observed, but the latter has not been identified in experiments~\cite{A33,A34}.

The Fermi energy is located in the energy region of the $4d$ band of Rh in CeRh$_{3}$, therefore the 
HI near the Fermi 
energy is strong, about three times greater than that of CePd$_{3}$.
The $4f$ band width of CeRh$_{3}$ is about twice that of CePd$_{3}$ in the LDA calculation.
In the DMFT calculation, the characteristic energy scales are drastically different from each other, 
300 and 5000 K.

In CeRh$_{3}$, the ARPES shows an appreciable DOS slightly density below the Fermi energy 
and is weakly dependent on the wave vector, although no flat $4f$ bands exist near the Fermi energy in the 
RNB dispersion.
The spectra have stronger intensity halfway along the $\Sigma$ axis from the $\Gamma$ to M points of the Brillouin zone.
This seems to be similar to the results of experiments, but further studies to separate the surface effects are 
necessary~\cite{A55}.

General features of the experimental results for CePd$_{3}$ and CeRh$_{3}$ are reproduced by the  
DMFT band calculation with the LMTO+NCA$f^{2}$vc scheme. 
However, 
the present DMFT calculation gives a higher Kondo temperature than that in the results of the detailed analysis 
of experiments.
Moreover, there is some arbitrariness in the choice of the target value of the $4f$ occupation number, $n_{f}({\rm rsl.target})$. 
According to calculations for AuCu$_{3}$-type Ce compounds, 
accurate calculated results seem to be obtained when a $4f$ occupation number between 90\% and 95\% of the LDA value is used. 
We have used 94\% in the present calculation, and we obtained 310~K~(27~meV) for $T_{\rm K}$ of CePd$_{3}$ 
($n_{f}({\rm rsl.target})=0.98$) and 
$ 4700$~K~(0.03~Ry) for $T_{\rm K}$ of CeRh$_{3}$ ($n_{f}({\rm rsl.target})=0.94$) 
from the magnetic excitation spectrum.
When we perform calculations using 90\%, the $T_{\rm K}$ are $400$~K 
for CePd$_{3}$ ($n_{f}({\rm rsl.target})=0.94$) and $ 6300$~K 
for CeRh$_{3}$ ($n_{f}({\rm rsl.target})=0.90$).
These values are not greatly different from the previous values because these compounds belong to a group of materials having 
high $T_ {\rm K} $.
For materials with lower $T_{\rm K}$, $T_{\rm K}$ drastically depends on the choice of $n_{f}({\rm rsl.target})$. 
Careful treatment of the target value is necessary in  such cases. 

The  effectiveness  and some of the weaknesses (for example, the correct calculation of the occupied electron number) of the present DMFT 
scheme are recognized.
Calculations of the $4f$ band state in CeIn$_{3}$ and CeSn$_{3}$, and also various Ce compounds will be 
carried out in the near future. 

At the end of this paper, we refer the very early and recent application of methods similar to NCA$f^{2}$vc to the 
DMFT band calculation.
L\ae gsgaard and Svane calculated the band structure of Ce pnictides in 1998~\cite{A57}.
Recently, Haule {\it et al.} studied the $\alpha \rightarrow \gamma$ transition of Ce metal~\cite{A58},
and Shim {\it et al.} studied the electronic band structure of CeIrIn$_{5}$~\cite{A59}.
The CFS of the self-energy was not considered in those studies.

\section*{Acknowledgments}

The author would like to thank  H. Shiba 
for encouragements, and  
Y. Kuramoto and J. Otsuki for important  comments on the resolvent
method, H. Harima for valuable comments on the band calculation method,
and Y. Shimizu for valuable collaboration in the early stage of developing the LMTO+NCA$f^{2}$vc code.
This work was partly supported by Grants-in-Aid for Scientific Research C (No. 21540372)
from JSPS, and   
on Innovative Areas ``Heavy Electrons''(No. 21102523) 
and for Specially Promoted Research (No. 18002008)
from MEXT.

\vfill\eject

\appendix

\section{Occupation number
}

The calculation based on the 
NCA$f^{2}$vc method usually does not satisfy the FL sum rule, i.e., the 
integral of the total DOS below the Fermi energy, 
$N({\rm total}; \rho)=\int{\rm d}\VEP\frac{1}{N}\sum_{\Bk}(-\frac{1}{\pi}\Im G(\VEP+{\rm i}\delta;\Bk))f(\VEP)$, 
is not equal to the value 
$N({\rm total};\ln G)=\int{\rm d}\VEP\frac{1}{N}\sum_{\Bk}(-\frac{1}{\pi}\Im\ln\det G(\VEP+{\rm i}\delta;\Bk))
(-\frac{\P f(\VEP)}{\P \VEP})$,
where $f(\VEP)$ is the Fermi distribution function at $T=0$.
In the RNB calculation, in which the self-energy term is approximated by 
$\Sigma_{\Gamma}(z)\approx\Re\Sigma_{\Gamma}(E_{\rm F})+\frac{\P\Re \Sigma_{\Gamma}(z)}{\P z}|_{E_{\rm F}}(z-E_{\rm F})$, 
we obtain real eigen energies $E_{\rm RNB}(\lambda\Bk)$ where $\lambda$ is the band suffix.
The number of occupied states in the RNB band, 
$N({\rm total};{\rm RNB})=\frac{1}{N}\sum_{\Bk}f(E_{\rm RNB}(\lambda\Bk))$, is expected to agree with 
$N({\rm total};\ln G)$ at $T=0$ if the imaginary part can be neglected near the Fermi energy.
However, the imaginary component has a considerable magnitude in NCA$f^{2}$vc even at very low temperatures; this is partly 
because $T$ must maintain the condition $T \gtrsim 0.1T_{\rm K}$.
For example, $N({\rm total};\rho)$, $N({\rm total};\ln G)$, and $N({\rm total};{\rm RNB})$ are 33.59, 34.16, and 34.01, 
respectively, at $T=18.75$ K in CePd$_{3}$.
The difference between these values becomes serious when the detailed structure of the Fermi surface is discussed.
In this study, we use $N({\rm total};{\rm RNB})$ to determine the Fermi energy $E_{\rm F}$ at $T=0$ because 
this quantity is directly related to the occupation number of electrons
calculated from the volume of the occupied states in the $\Bk$ space.

\section{Cauchy integral of using spline interpolation
}
The Cauchy integral appears in various places in the DMFT calculation. 
Therefore, an efficient and accurate numerical calculation of the Cauchy integral is needed.
We briefly explain a method of using the spline interpolation for the DOS. 
Let us assume that numerical data of DOS $\{y_{i}\}$ at energy points $\{x_{i}\}$ ($i=1, N_{\rm data number}$) are given.
In the cubic spline interpolation~\cite{A60}, the DOS in the interval $[x_{i},x_{i+1}]$ is 
expressed as
\Beqa
\rho_{3,i}(x) =\frac{1}{6h_{i}}\{(x_{i+1}-x)^{3}M_{i}+(x-x_{i})^{3}M_{i+1}\} \nonumber \\
      +(y_{i}  -\frac{h_{i}^{2}M_{i}  }{6})\frac{x_{i+1}-x}{h_{i}} 
      +(y_{i+1}-\frac{h_{i}^{2}M_{i+1}}{6})\frac{x  -x_{i}}{h_{i}},
\Eeqa
where $h_{i}=x_{i+1}-x_{i}$.
The quantity $M_{i}$ is the  second derivative of the DOS at $x=x_{i}$, and is given by the usual 
procedure of the spline interpolation.

The integral of the interval $[x_{i},x_{i+1}]$ is calculated as
\Beqa
\int_{x_{i}}^{x_{i+1}}\frac{\rho_{3,i}(x)}{z-x}{\rm d}x
 = (x_{i+1}-z)^{2}\frac{M_{i}       }{6} -(z-x_{i})^{2}\frac{M_{i+1}       }{6}   \nonumber \\
  +(x_{i+1}-z)    \frac{M_{i}h_{i}}{12}-(z-x_{i})     \frac{M_{i+1}h_{i}}{12} \hspace{1cm} \nonumber \\
  +(y_{i}  -\frac{h_{i}^{2}M_{i}  }{9}) 
   -(y_{i+1}-\frac{h_{i}^{2}M_{i+1}}{9}) \hspace{1cm} \nonumber \\ 
   -\rho_{3,i}(z)\ln\frac{z-x_{i+1}}{z-x_{i}}. \hspace{2cm}
\label{B2}
\Eeqa
The total integral is given by summing the contribution from each interval.
Equation (\ref{B2}) is expressed as a combination of powers of quantities $(x_{i+1}-z)$ and 
$(z-x_{i})$; therefore it is applicable for $z$ when $|z-\frac{x_{i}+x_{i+1}}{2}|$ is not much
lager than $h_{i}$.
When $z$ approaches the edge of the integration, i.e., $z \rightarrow x_{i}$ or $x_{i+1}$, 
the singularity of the logarithm term is removed by the counter contribution of  
the neighboring $[x_{i-1}, x_{i}]$ or $[x_{i+1},x_{i+2}]$ region.
The round-off error due to the subtraction of logarithm terms of neighboring regions is not serious 
even when $z$ is extremely near
the edge point, because the divergence of the logarithm is very weak.
The integral (\ref{B2}) is expected to be $O(z-\frac{x_{i}+x_{i+1}}{2})^{-1}$
when $|z-\frac{x_{i}+x_{i+1}}{2}| \gg h_{i}$.
Therefore, mutual cancellation occurs among terms in (\ref{B2}) in this limit. 
We find that the integral is re-expressed  by a compact form in this case 
\Beqa
\int_{x_{i}}^{x_{i+1}}\frac{\rho_{3,i}(x)}{z-x}{\rm d}x \hspace{4.5cm} \nonumber \\
= -\sum\limits_{\nu=1}^{\infty}[(\frac{h_{i}}{x_{i+1}-z})^{\nu}(\frac{y_{i}  }{\nu+1}-\frac{h_{i}^{2}M_{i}  }{3(\nu+3)(\nu+1)}) \nonumber \\
                        -(\frac{h_{i}}{z-  x_{i}})^{\nu}(\frac{y_{i+1}}{\nu+1}-\frac{h_{i}^{2}M_{i+1}}{3(\nu+3)(\nu+1)})]. 
\Eeqa
This equation gives a highly accurate estimation of the integral even when terms are truncated up to $\nu \sim 10$.

In most cases, it is convenient to use the linear interpolation scheme with a very fine mesh for the DOS, 
\Beqa
\rho_{1,i}(x) =m_{i}(x-\frac{x_{i}+x_{i+1}}{2})+\frac{y_{i}+y_{i+1}}{2},
\Eeqa
where $m_{i}=\frac{y_{i+1}-y_{i}}{h_{i}}$.
The Cauchy integral is expressed by the following forms  
\Beqa
\int_{x_{i}}^{x_{i+1}}\frac{\rho_{1,i}(x)}{z-x}{\rm d}x 
 = -m_{i}h_{i}-\rho_{1,i}(z)\ln\frac{z-x_{i+1}}{z-x_{i}},
\Eeqa
and for $|z-\frac{x_{i}+x_{i+1}}{2}| \gg h_{i}$, 
\Beqa
\int_{x_{i}}^{x_{i+1}}\frac{\rho_{1,i}(x)}{z-x}{\rm d}x 
= &-\frac{1}{2}\sum\limits_{\nu=1}^{\infty}[
  (\frac{h_{i}}{x_{i+1}-z})^{\nu}(\frac{y_{i+1}}{\nu}-\frac{m_{i}h_{i}}{\nu+1}) \nonumber \\
  &- (\frac{h_{i}}{z-  x_{i}})^{\nu}(\frac{y_{i}  }{\nu}+\frac{m_{i}h_{i}}{\nu+1}) ].
\Eeqa


\begin{thebibliography}{9}
\bibitem{A41}
The summation over $\Bk$ is carried out on a mesh dividing the $\Gamma-X$ axis into 8 parts.
Usually this gives a nearly smooth spectrum for $4f$ components, but gives fictitious fine structures for non-$4f$ 
components.
The total DOS is calculated by smoothing with the introduction of the imaginary part $\gamma=0.01$~Ry 
in the energy variable.
\bibitem{D2}
In NCA$f^{2}$vc, the atomic energy levels are treated as independent input parameters. The energy level of the $f^{0}$ state is $E(f^{0}) =0$ 
and the energy level of the $f^{1}$ state is $E(f^{1},\Gamma)=\VEP_{\Gamma}$ when levels are measured from $E_{\rm F}$. 
The energy level of the $f^{2}$ state is assumed to be  
$E(f^{2}) = 2\VEP(f)+U$, where $\VEP(f)=\frac{1}{6}(2\VEP_{\Gamma_{7}}+4\VEP_{\Gamma_{8}})$ is the 
average of $f$ levels of the $j=5/2$ multiplet, and $U$ is the Coulomb constant.
\bibitem{D1}
The impurity $4f$ Green function is given using  the $4f$ DOS $\rho^{\rm (imp.)}_{\Gamma}(x)$ as the Cauchy integral: 
$G_{\Gamma}^{\rm (imp.)}(z)=\int{\rm d}x\frac{\rho^{\rm (imp.)}_{\Gamma}(x)}{z-x}$.
It is also expressed as  
$G_{\Gamma}^{\rm (imp.)}(z)=1/[z-\VEP^{\rm a}_{\Gamma}-\Sigma_{\Gamma}(z)-\Sigma_{\Gamma}^{\rm (h)}(z)]$, 
where $\VEP^{\rm a}_{\Gamma}$ is the impurity level, and $\Sigma_{\Gamma}(z)$ and  $\Sigma^{\rm (h)}_{\Gamma}(z)$ 
are, respectively,
the self-energies due to the Coulomb interaction and the hybridization~\cite{A6}. 
The self-energy $\Sigma_{\Gamma}(z)$ is calculated as 
$\Sigma_{\Gamma}(z)=z-\VEP^{\rm a}_{\Gamma}-G^{\rm (imp.)}_{\Gamma}(z)^{-1}-\Sigma^{\rm (h)}_{\Gamma}(z)$
by solving the impurity model with the HI $-\Im\Sigma_{\Gamma}^{\rm (h)}(x+\I\delta)$. 
The self-consistent condition on the $4f$ DOS in DMFT is $\rho^{\rm (imp.)}_{\Gamma}(x)=\rho^{(\rm band)}_{\Gamma}(x)$. 
Under the self-consistent condition, 
the self-energy due to the hybridization is given by
$\Sigma^{\rm (h)}_{\Gamma}(z)=z-\VEP^{\rm a}_{\Gamma}-G^{\rm (imp.)}_{\Gamma}(z)^{-1}-\Sigma_{\Gamma}(z)$.
\bibitem{A44}
We find that the maximum relative difference of the $4f$~DOS hardly decreases below 0.5, even when we 
advance the iteration process.
The average of the relative differences of the Greenian on the imaginary axis is monitored as the convergence parameter, 
but this also hardly decreases below $0.3 \times 10^{-2}$.
We usually stop the iteration process when this parameter reaches approximately $0.3 \times 10^{-2}$ and 
shows stationary behavior.
\bibitem{A49}
When the CFS of the electrostatic term is not included, a large number of iteration loops are needed to reach 
a solution for large CFS because the calculation starts with a condition of equal energy levels: 
$\VEP^{\rm a}_{\Gamma 7}=\VEP^{\rm a}_{\Gamma 8}$.
\bibitem{A50}
When we perform the LDA calculation for a fictitious LaPd$_{3}$ with the lattice constant of CePd$_{3}$, 
the occupation number on the $4f$ state is estimated to be 0.188, which is larger than that of usual 
compounds, about 0.1.
This large occupation is induced through the strong hybridization between the $4f$ state and the $4d$(Pd) state 
located in the energy region deep below $E_{\rm F}$.
Some parts of the $4f$ occupation in LaPd$_{3}$ may be ascribed to the 
tail of the $4d$(Pd) extending to the 
Ce region.
\bibitem{D3}
Under the self-consistent condition, 
the self-energy due to the hybridization is given by
$\Sigma^{\rm (h)}_{\Gamma}(z)=z-\VEP^{\rm a}_{\Gamma}-G^{\rm (imp.)}_{\Gamma}(z)^{-1}-\Sigma_{\Gamma}(z)$,
where $\VEP^{\rm a}_{\Gamma}$ is the impurity level, 
$G^{\rm (imp.)}_{\Gamma}(z)$ is the impurity Green's function and  
$\Sigma_{\Gamma}(z)$ is the self-energy due to the Coulomb interaction.
The HI shown in Fig.~\ref{fig:CePd3-38-mix}  is defined as $-\Im\Sigma_{\Gamma}^{\rm (h)}(x+\I\delta)$.
This is the quantity corresponding to $\sum_{\nu\Bk}\pi|v_{\Gamma}(\nu\Bk)|^{2}\delta(x-E_{\nu\Bk})$ in the usual impurity problem, 
where $v_{\Gamma}(\nu\Bk)$ is the $c-f$ hybridization matrix and $E_{\nu\Bk}$ is the energy of the band state.
\bibitem{D4}
The $4f$ electron occupation number calculated using the RNB is 0.84 at $T=37.5$ K for CePd$_{3}$.
This value is large compared with the value supposed from the impression of Fig.~\ref{fig:CePd3-rnb} in which  
only the bottoms of $4f$ bands are occupied.
The $4f$ occupation number through the hybridization with conduction bands is considerable in the RNB picture of CePd$_{3}$.
The LDA calculation also gives $4f$ bands where only their bottoms are occupied in the same way as in the RNB calculation, 
and leads to the $4f$ occupation number of 1.045.
\bibitem{D5}
The $4f$ DOS calculated by  DMFT is shown in Fig.~\ref{fig:CePd3-38-flsp}.
The spectrum intensity in the deep region from $E_{\rm F}$ (``the incoherent component'' in the terminology of band theory and ``the $f^{0}$ excitation 
part'' in the terminology of the local Kondo problem) has a considerably high intensity.
However, this deep energy component is disregarded in RNB, but 
its intensity is recovered  as the full value of the intensity of the ``coherent part'' in the calculation of the occupied state.
Only the band dispersion in the  neighborhood of $E_{\rm F}$ and the count of the occupied $\Bk$ states at very low temperatures compared with 
the Kondo temperature have  meaning in the RNB calculation.
\bibitem{D6}
Note that the temperature dependence of $N({\rm total};\rho)$ is not strong: it is 33.59 at $T=18.75$~K, and is 33.60 at $T=300$~K.
\end{thebibliography}
\end{document}